# A stress-driven local-nonlocal mixture model for Timoshenko nano-beams

Raffaele Barretta[a], Andrea Caporale[b], S. Ali Faghidian[c], Raimondo Luciano[b], Francesco Marotti de Sciarra[a], Carlo Maria Medaglia[d]

[a] *Department of Structures for Engineering and Architecture, University of Naples Federico II, e-mails: rabarret@unina.it - marotti@unina.it*
[b] *Department of Civil and Mechanical Engineering, University of Cassino and Southern Lazio, e-mail: a.caporale@unicas.it - luciano@unicas.it*
[c] *Department of Mechanical Engineering, Science and Research Branch, Islamic Azad University, Tehran, Iran, e-mail: faghidian@gmail.com*
[d] *Link Campus University, e-mail: c.medaglia@unilink.it*

---

## Abstract

A well-posed stress-driven mixture is proposed for Timoshenko nano-beams. The model is a convex combination of local and nonlocal phases and circumvents some problems of ill-posedness emerged in strain-driven Eringen-like formulations for structures of nanotechnological interest. The nonlocal part of the mixture is the integral convolution between stress field and a bi-exponential averaging kernel function characterized by a scale parameter. The stress-driven mixture is equivalent to a differential problem equipped with constitutive boundary conditions involving bending and shear fields. Closed-form solutions of Timoshenko nano-beams for selected boundary and loading conditions are established by an effective analytical strategy. The numerical results exhibit a stiffening behavior in terms of scale parameter.

---

## 1. Introduction

The local continuum theory fails to capture size effects in nanodevices, such as Nano-Electro-Mechanical Systems (NEMS) [1, 2, 3], and is not able to describe the behavior of structures characterized by an external overall length not much greater than the material internal characteristic length. The scientific literature suggests various approaches (atomistic, strain gradient, Eringen nonlocal theory and so on) to be considered when the local continuum theory is inadequate, see e.g. [4, 5, 6, 7, 8, 9, 10] and references cited therein. In this paper, an innovative stress-driven nonlocal mixture is proposed for Timoshenko nano-beams. Such a model is combination of local and nonlocal elastic phases and differs from Eringen-like [11, 12, 13, 14], strain gradient [15, 16, 17, 18, 19], couple stress models [20, 21] and other approaches available in literature [22, 23, 24, 25, 26, 27, 28, 29, 30]. The need of the proposed stress-driven approach with mixture of local and nonlocal phases derives from a recent research [31] on behavior of the strain-driven nonlocal model for bounded nanostructures, where the bending field is expressed as convolution of elastic curvature with an averaging kernel assuming an exponential expression. The associated elastostatic problem can be solved provided that the bending moment field satisfies suitable constitutive boundary conditions (CBC). This verification usually fails in cases of applicative interest, arising an ill-posedness of nonlocal nonlocal problems in nanomechanics. In order to overcome this difficulty, Romano and Barretta [32, 33] proposed a stress-driven nonlocal integral model, where the bending field is placed in the proper position of input variable so that

## 1. Introduction

The local continuum theory fails to capture size effects in nanodevices, such as Nano-Electro-Mechanical Systems (NEMS) [1, 2, 3], and is not able to describe the behavior of structures characterized by an external overall length not much greater than the material internal characteristic length. The scientific literature suggests various approaches (atomistic, strain gradient, Eringen nonlocal theory and so on) to be considered when the local continuum theory is inadequate, see e.g. [4, 5, 6, 7, 8, 9, 10] and references cited therein. In this paper, an innovative stress-driven nonlocal mixture is proposed for Timoshenko nano-beams. Such a model is combination of local and nonlocal elastic phases and differs from Eringen-like [11, 12, 13, 14], strain gradient [15, 16, 17, 18, 19], couple stress models [20, 21] and other approaches available in literature [22, 23, 24, 25, 26, 27, 28, 29, 30]. The need of the proposed stress-driven approach with mixture of local and nonlocal phases derives from a recent research [31] on behavior of the strain-driven nonlocal model for bounded nanostructures, where the bending field is expressed as convolution of elastic curvature with an averaging kernel assuming an exponential expression. The associated elastostatic problem can be solved provided that the bending moment field satisfies suitable constitutive boundary conditions (CBC). This verification usually fails in cases of applicative interest, arising an ill-posedness of nonlocal nonlocal problems in nanomechanics. In order to overcome this difficulty, Romano and Barretta [32, 33] proposed a stress-driven nonlocal integral model, where the bending field is placed in the proper position of input variable so that the constitutive law is evaluated by convolution between bending field and an averaging kernel. Strain-driven mixtures of local and nonlocal material laws are also considered in literature [34, 35, 36, 37, 38, 39, 40]: the local elastic fraction of the mixture induces well-posedness [31]. However, this beneficial effect does not hold for vanishing local fractions, see e.g. [41]. On the contrary, the stress-driven theory does not suffer the limiting behavior of strain-driven formulations as the local fraction tends to zero. The stress-driven model has been adopted in various problems such as bending of functionally graded nano-beams [42, 43] and nonlocal thermoelastic behavior of nano-beams [44]. Such a model provides a stiffening structural behavior in accordance with experimental evidences [15, 45].

In this paper, a stress-driven local-nonlocal mixture defined by convexly combining local and nonlocal phases is adopted and presented in Section 2. In Section 3, it is assumed that the kernel function appearing in the integral convolution is a bi-exponential function. Under this assumption, the integral formulation in Section 2 is equivalent to a differential problem equipped with suitable constitutive boundary conditions. In Sections 4.1 and 4.2, procedures for obtaining closed-form solutions of Timoshenko nano-beams are illustrated by applying integral and differential formulations. Such procedures provide general closed-form solutions, which are valid for any kind of boundary conditions and external loads applied to the nano-beam and are evaluated for some fundamental schemes in Appendix A. Closed-form solutions may have long expressions because contain the effects of both the flexural and shear displacements. For this reason, numerical solutions have been presented in Section 5 for selected stubby nano-beams.

## 2. Mixture stress-driven integral model for Timoshenko beams

The mixture stress-driven integral model (MStreDM) is applied to Timoshenko nano-beams. MStreDM is the generalization to stubby beams of a previous constitutive version formulated in [42] for Bernoulli-Euler nano-beams in which the flexural curvature field $\chi(x)$ is expressed by the following two-phase law:

$$\chi(x) = \alpha \frac{M(x)}{EI} + (1-\alpha) \int_0^L \psi(x-t, L_c) \frac{M(t)}{EI} dt, \qquad (1)$$

with $L > 0$ nano-beam length; $EI$ and $M$ stand for local bending stiffness and moment, respectively; $\psi$ is an averaging kernel dependent on the small scale coefficient $L_c > 0$. In the proposed MstreDM for Timoshenko beams, the shear deformation is convex combination of local and nonlocal phases, with the nonlocal phase given by Romano-Barretta stress-driven integral law. The shear deformation, evaluated as difference between the slope $\theta(x)$ of the beam deformed center-line and the rotation $\phi(x)$, is expressed by:

$$\theta(x) - \phi(x) = \alpha \frac{T(x)}{\kappa G A} + (1-\alpha) \int_0^L \psi(x-t, L_c) \frac{T(t)}{\kappa G A} dt, \qquad (2)$$

with $GA$ local shear stiffness and $\kappa$ shear factor. In (1) and (2), bending moment $M$ and shear force $T$ are equilibrated. The phase parameter $\alpha$ belongs to $[0,1]$. Integral equations (1) and (2) are coupled through the kinematic condition $\chi(x) = \phi'(x)$ and the static condition $M'(x) = -T(x)$. Figure 1 shows the meaning of some kinematic variables involved in the adopted nano-beam model.

### 3. Mixture stress-driven differential law for Timoshenko beams

Assuming the following exponential expression for the function $\psi$:

$$\psi(x, L_c) = \frac{1}{2L_c} \exp\left(-\frac{|x|}{L_c}\right),$$ (3)

it will be proven that the problem defined by integral equations (1) and (2) admits an equivalent differential form subject to constitutive boundary conditions. Barretta et al. [42] have demonstrated that the nonlocal constitutive law (1) is equivalent to the following second-order differential equation when $\psi(x, L_c)$ is assumed equal to the Helmholtz kernel (3):

$$\chi''(x) - \frac{1}{L_c^2}\chi(x) = \alpha\frac{M''(x)}{EI} - \frac{1}{L_c^2}\frac{M(x)}{EI},$$ (4)

where $x \in [0, L]$ and bending curvature $\chi(x)$ satisfies the following constitutive boundary conditions (CBCs)

$$\begin{cases} \chi'(0) - \dfrac{1}{L_c}\chi(0) = \alpha\dfrac{M'(0)}{EI} - \dfrac{\alpha}{L_c}\dfrac{M(0)}{EI}, \\[2mm] \chi'(L) + \dfrac{1}{L_c}\chi(L) = \alpha\dfrac{M'(L)}{EI} + \dfrac{\alpha}{L_c}\dfrac{M(L)}{EI}. \end{cases}$$ (5)

In (5), derivatives are evaluated at the end points of the nano-beam, such as $\chi'(0) \equiv \frac{d\chi}{dx}|_{x=0}$. Next, we demonstrate that the non-local constitutive law (2) is also equivalent to a second-order differential equation subject to constitutive boundary conditions. Let us split the integral in (2) by setting

$$\theta(x) - \phi(x) = \alpha\frac{T(x)}{\kappa GA} + (1-\alpha)(\gamma_1(x) + \gamma_2(x))$$ (6)

with

$$\begin{cases} \gamma_1(x) = \int_0^x \psi(x-t, L_c) \dfrac{T(t)}{\kappa GA} dt = \int_0^x \dfrac{1}{2L_c} \exp\left(\dfrac{t-x}{L_c}\right) \dfrac{T(t)}{\kappa GA} dt, \\[4mm] \gamma_2(x) = \int_x^L \psi(x-t, L_c) \dfrac{T(t)}{\kappa GA} dt = \int_x^L \dfrac{1}{2L_c} \exp\left(\dfrac{x-t}{L_c}\right) \dfrac{T(t)}{\kappa GA} dt. \end{cases} \tag{7}$$

In (7), the dependence of $\gamma_1$ and $\gamma_2$ on $L_c$ is omitted for lightening the expressions in the next equations. Taking the first derivative, we get

$$\begin{cases} \gamma_1'(x) = \dfrac{1}{L_c} \left( +\dfrac{T(x)}{2\kappa GA} - \gamma_1(x) \right), \\[4mm] \gamma_2'(x) = \dfrac{1}{L_c} \left( -\dfrac{T(x)}{2\kappa GA} + \gamma_2(x) \right) \end{cases} \tag{8}$$

and, from (8), differentiating (6) yields the relation:

$$\theta'(x) - \phi'(x) = \alpha \frac{T'(x)}{\kappa GA} + \frac{1-\alpha}{L_c}(\gamma_2(x) - \gamma_1(x)). \tag{9}$$

Differentiating again, from (8) we get

$$\theta''(x) - \phi''(x) = \alpha \frac{T''(x)}{\kappa GA} + \frac{1-\alpha}{L_c^2} \left( -\frac{T(x)}{\kappa GA} + \gamma_1(x) + \gamma_2(x) \right), \tag{10}$$

which can be rewritten as

$$\begin{aligned} \theta''(x) - \phi''(x) = \\ \alpha \frac{T''(x)}{\kappa GA} + \frac{1-\alpha}{L_c^2} \left( -\frac{T(x)}{\kappa GA} \right) + \frac{1}{L_c^2} \left( \theta(x) - \phi(x) - \alpha \frac{T(x)}{\kappa GA} \right) \end{aligned} \tag{11}$$

or, more concisely, as

$$\theta''(x) - \frac{\theta(x)}{L_c^2} = \alpha \frac{T''(x)}{\kappa GA} - \frac{1}{L_c^2} \frac{T(x)}{\kappa GA} + \phi''(x) - \frac{\phi(x)}{L_c^2}. \tag{12}$$

Eqs. (2) and (12) provide the Bernoulli-Euler kinematic relation $\phi(x) = \theta(x)$ when the shear stiffness $GA$ is infinitely large. We get the constitutive boundary conditions for (12) by evaluating (6) at the boundary points $x = 0$ and $x = L$ of the nano-beam axis:

$$(1 - \alpha)\, \gamma_2(0) = \theta(0) - \phi(0) - \alpha \frac{T(0)}{\kappa GA},$$
$$(1 - \alpha)\, \gamma_1(L) = \theta(L) - \phi(L) - \alpha \frac{T(L)}{\kappa GA} \tag{13}$$

and then by substituting (13) in (9) evaluated at the same boundary points:

$$\theta'(0) - \frac{1}{L_c}\theta(0) = \alpha \frac{T'(0)}{\kappa GA} - \frac{\alpha}{L_c}\frac{T(0)}{\kappa GA} + \phi'(0) - \frac{1}{L_c}\phi(0),$$
$$\theta'(L) + \frac{1}{L_c}\theta(L) = \alpha \frac{T'(L)}{\kappa GA} + \frac{\alpha}{L_c}\frac{T(L)}{\kappa GA} + \phi'(L) + \frac{1}{L_c}\phi(L). \tag{14}$$

Concluding, the integral formulation governed by equations (1) and (2) admits the differential formulation represented by the differential equations (4) and (12) subject to the constitutive boundary conditions (5) and (14), respectively. The integral formulation provides the same solution given by the differential formulation.

## 4. Nonlocal solution procedures

Next, we describe the procedures for obtaining the solution of the Timoshenko nano-beams by applying both the integral formulation explained in Section 2 and the the differential formulation explained in Section 3. These procedures provide general solutions valid for any kinds of boundary conditions and external loads applied to the nano-beam. Specifically, the considered external loads are a transverse distributed load $q_y(x)$, a concentrated load $F$ or a concentrated couple $m$, but other load such as distributed couples may also be introduced in the formulation. The procedure based on the integral formulation is explained in the following Section 4.1 whereas the procedure based on the differential form is described in Section 4.2.

### 4.1. Integral procedure

Solving the equilibrium differential equation $M''(x) = q_y(x)$, the bending moment is given by

$$M(x) = \int_0^x (x-s)q_y(s)ds + A_1 x + A_2,$$ (15)

resulting an expression in terms of the two integration constants $A_1$ and $A_2$. Setting

$$f(x) = \frac{1}{EI}\int_0^x (x-s)q_y(s)ds + A_1 x + A_2,$$ (16)

and taking into account (1) and (15), the bending curvature $\chi(x)$ is

$$\chi(x) = \alpha f(x) + (1-\alpha)\int_0^L \psi(x-t, L_c)\, f(t)dt.$$ (17)

The evaluation of (17) provides the curvature $\chi(x)$ again in terms of the two integration constants $A_1$ and $A_2$. Then, solving the kinematic compatibility differential equation $\chi(x) = \phi'(x)$, the cross-sectional rotation is

$$\phi(x) = \int_0^x \chi(s)ds + A_3,\tag{18}$$

which, by virtue of (17), results an expression in terms of the three integration constants $A_1$, $A_2$ and $A_3$. Solving the equilibrium differential equation $M'(x) = -T(x)$ and taking into account (15), the shear force is given by

$$T(x) = -\int_0^x q_y(s)ds - A_1.\tag{19}$$

Setting

$$h(x) = -\frac{1}{\kappa GA}\left(\int_0^x q_y(s)ds + A_1\right),\tag{20}$$

and taking into account (2) and (19), the slope $\theta(x)$ of the transverse displacements is

$$\theta(x) - \phi(x) = \alpha h(x) + (1-\alpha)\int_0^L \psi(x-t, L_c)\,h(t)dt,\tag{21}$$

which now is evaluated with rotation $\phi(x)$ given by (18), providing a new expression of $\theta(x)$. Solving the kinematic compatibility differential equation $\theta(x) = v'(x)$ with $\theta(x)$ given by (21), the transverse displacement is

$$v(x) = \int_0^x \theta(s)ds + A_4,\tag{22}$$

resulting an expression in terms of the four integration constants $A_1$, $A_2$, $A_3$ and $A_4$ to be determined by using the canonical boundary conditions. There-fore, the last step is determining the above-mentioned integration constants by solving a system of four linear boundary conditions in the unknowns $A_1$, $A_2$, $A_3$ and $A_4$. These boundary conditions are obtained by imposing that displacement $v(x)$, rotation $\phi(x)$, shear $T(x)$ and/or bending moment $M(x)$ assume suitable values at the boundary points $x = 0$ and $x = L$ of the nano-beam. Observing the structure of the previous equations, it is also clear that the integration constants have the following meaning:

$$A_1 = -T(0), \quad A_2 = M(0), \quad A_3 = \phi(0), \quad A_4 = v(0). \tag{23}$$

### 4.2. Differential procedure

In this section, the solution of the Timoshenko nano-beam is obtained by applying the differential formulation in Section 3, which involves the solution of two differential equations with the corresponding constitutive boundary conditions. The first step is solving the second-order differential equation (4) subject to the CBCs (5); taking into account (15) and (16), this task is equivalent to solve the following differential equation

$$\chi''(x) - \frac{1}{L_c^2}\chi(x) = \alpha f''(x) - \frac{1}{L_c^2}f(x) \tag{24}$$

with $x \in [0, L]$, subject to the following CBCs

$$\begin{cases} \chi'(0) - \dfrac{1}{L_c}\chi(0) = \alpha f'(0) - \dfrac{\alpha}{L_c}f(0), \\[3mm] \chi'(L) + \dfrac{1}{L_c}\chi(L) = \alpha f'(L) + \dfrac{\alpha}{L_c}f(L). \end{cases} \tag{25}$$

The solution $\chi(x)$ of (24) is used to evaluate the cross-sectional rotation

$$\phi(x) = \int_0^x \chi(s)ds + A_3, \tag{26}$$

which results an expression in terms of the three integration constants $A_1$, $A_2$ and $A_3$. The second step is solving the second-order differential equation (12) subject to the CBCs (14); taking into account (19) and (20), this task is equivalent to solve the following differential equation

$$\theta''(x) - \frac{\theta(x)}{L_c^2} = \alpha h''(x) - \frac{h(x)}{L_c^2} + \phi''(x) - \frac{\phi(x)}{L_c^2}, \tag{27}$$

with $\phi(x)$ given by (26), subject to the following CBCs

$$\begin{cases} \theta'(0) - \dfrac{1}{L_c}\theta(0) = \alpha h'(0) - \dfrac{\alpha}{L_c}h(0) + \phi'(0) - \dfrac{1}{L_c}\phi(0), \\[3mm] \theta'(L) + \dfrac{1}{L_c}\theta(L) = \alpha h'(L) + \dfrac{\alpha}{L_c}h(L) + \phi'(L) + \dfrac{1}{L_c}\phi(L). \end{cases} \tag{28}$$

Finally, the solution $\theta(x)$ of (27) is used to evaluate the displacement

$$v(x) = \int_0^x \theta(s)ds + A_4, \tag{29}$$

which is expressed in terms of the integration constants $A_1$, $A_2$, $A_3$ and $A_4$ to be determined by using the boundary conditions described in Section 4.1.

## 5. Solutions of Timoshenko nano-beams

The procedures described in Sections 4.1 and 4.2 provide closed-form solutions for kinematic and static entities (such as $v(x)$, $\phi(x)$, $T(x)$, etc.) of the Timoshenko nano-beam problem. Closed-form solutions $v(x)$ and $\theta(x)$ have long expressions because contain the effects of both the flexural and shear deformations. This leads to represent the solution of the Timoshenko nano-beams by means of graphics in this section, avoiding the writing of long symbolic expressions for $v(x)$, $\theta(x)$, etc. Closed-form solutions for some fundamental problems are obtained with the proposed methods and are reported in Appendix A. The following geometric dimensionless variables are introduced:

$$\xi = \frac{x}{L}, \quad \lambda = \frac{L_c}{L}. \tag{30}$$

Integral equation (1) shows that kinematic variable $\chi(x)$ is proportional to the external load contained in $M(x)$ and inversely proportional to the bending stiffness $EI$. Analogously from (2), kinematic variable $\gamma(x) = \phi(x) - \theta(x)$ is proportional to the external load contained in $T(x)$ and inversely proportional to the shear stiffness $GA$. This allows adopting dimensionless kinematic variables for representing the solution of nano-beams. To this end, the shear stiffness $GA$ is expressed in terms of the flexural stiffness $EI$ by introducing the dimensionless ratio

$$\beta = \frac{\kappa G A \, L^2}{EI}. \tag{31}$$

The expression of the dimensionless transverse displacement $v^*(\xi)$ used to graphically represent the solution depends on the kind of load applied to the nano-beam and is given by

$$
v^*(\xi) = \begin{cases} v(x)\dfrac{EI}{mL^2} & \text{if a concentrated couple } m \text{ is applied,} \\[2em] v(x)\dfrac{EI}{FL^3} & \text{if a transverse concentrated force } F \text{ is applied,} \\[2em] v(x)\dfrac{EI}{qL^4} & \text{if a transverse uniform load } q_y(x) = q \text{ is applied.} \end{cases}
$$

(32)

The dimensionless displacements defined in (32) are usually adopted for beams subject to pure flexural deflection without shear deformation; but substituting $\kappa GA = \beta \frac{EI}{L^2}$ in the closed-form expressions of $v(x)$, according to (31), displacements $v^*(\xi)$ defined in (32) also result dimensionless for beams exhibiting shear deformations in conformity with the Timoshenko model.

In the proposed nonlocal theory, the parameter $\lambda$ plays an important role. Setting $\lambda = 0$ provides the classical local Timoshenko beam solution [1], which can also be recovered by solving the problem with $\alpha = 1$, whatever value of $\lambda$ is adopted. Next, $v^*_{LT}(\xi)$ denotes the dimensionless transverse displacement of the local Timoshenko beam (LT). The nonlocal Timoshenko beam (NLT) occurs when $\lambda > 0$ and $\alpha \neq 1$. Another significant solution is obtained by setting $\lambda \to +\infty$: the corresponding dimensionless displacement $v^*(\xi)$ is denoted by $v^*_{T\infty}(\xi)$; from (1) and (2), it results $v^*_{T\infty}(\xi) = \alpha v^*_{LT}(\xi)$, see also Appendix A.

---

[1] In this paper, setting a parameter equal to zero (or $+\infty$) also means evaluating a limit, i.e. finding the solution as that parameter tends to zero (or $+\infty$). See also Appendix A.

The stiffness ratio $\beta = \frac{\kappa G A L^2}{EI}$ influences the magnitude of the flexural and shear deformations. As $\beta$ tends to $+\infty$ the shear deformation tends to vanish with respect to the flexural deformation. Therefore, setting $\beta = +\infty$ provides the Bernoulli-Euler beam, which is local (LBE) if $\lambda = 0$ and is nonlocal (NLBE) if $\lambda > 0$ and $\alpha \neq 1$. The mixture stress-driven formulations for Bernoulli-Euler nano-beams has been treated in greater detail in [42]. Analogously to the Timoshenko beam, $v^*_{LBE}(\xi)$ denotes the dimensionless transverse displacement of the local Bernoulli-Euler beam and $v^*_{BE\infty}(\xi) = \alpha v^*_{LBE}(\xi)$ denotes the dimensionless transverse displacement obtained by setting $\lambda = +\infty$ for the NLBE.

Figure 2 shows dimensionless displacement $v^*$ against $\xi$ for a cantilever subject to a transverse concentrated load at the free end point. The curves of the figure corresponding to the local and nonlocal Timoshenko beams are denoted by LT and NLT, respectively. In the same figure, the curves corresponding to local and nonlocal Bernoulli-Euler beams are also reported and denoted by LBE and NLBE, respectively. For nonlocal beams, the mixture parameter $\alpha$ is assumed equal to 0.5. For the Timoshenko beams, the stiffness ratio $\beta$ is assumed equal to 4. The comparison between Timoshenko and Bernoulli-Euler beams in Figure 2 highlights the increment of displacement $v^*(\xi)$ due to the shear deformation effect characterizing the Timoshenko model. The curves of the figure refer to several values of $\lambda$ varying in the set $\mathcal{S} = \{0, 0.1, 0.2, 0.3, 0.4, 0.5, +\infty\}$. The parameter $\lambda$ has the effect of reducing the displacement $v^*(\xi)$, that is a larger $\lambda$ involves a smaller displacement $v^*(\xi)$ for a given value of $\alpha$. Therefore, $v^*_{T\infty}(\xi)$ and $v^*_{LT}(\xi)$ $\forall \xi \in [0, 1]$ are lower and upper bounds, respectively, for the Timo-shenko dimensionless displacement $v^*(\xi)$ corresponding to any $\lambda \in \mathcal{S}$. An analogous observation holds for the Bernoulli-Euler beams.

In the proposed mixture formulation, the transition from local to nonlocal models is governed by the parameter $\alpha$. As the nonlocal model is stiffer than the local model, the parameter $\alpha$ has the effect of increasing the displacement $v^*(\xi)$, that is a larger $\alpha$ involves a larger displacement $v^*(\xi)$ for a given value of $\lambda$. Figures 3-10 show dimensionless displacement $v^*$ against $\xi$ for different kinds of loads and boundary conditions. Table 1 shows the boundary conditions of the generic nano-beam considered in Figures 2-10: specifically, the first column of the table reports the name of the nano-beam and the figure where the nano-beam is investigated; the second column of the table reports the corresponding boundary conditions. The transverse distributed load is absent except in nano-beams $C$-$q$, $SS$-$q$, $CS$-$q$ and $CC$-$q$, where a uniformly transverse distributed load $q$ is applied. In Figures 3-10, the curves corresponding to the local and nonlocal Timoshenko beams are denoted by LT and NLT, respectively. In the same figures, the curves corresponding to local and nonlocal Bernoulli-Euler beams are denoted by LBE and NLBE, respectively. Moreover, the mixture parameter $\alpha$ is assumed equal to 0.5 for nonlocal beams and the stiffness ratio $\beta$ is assumed equal to 4 for the Timoshenko beams. The considerations made for Figure 2 are also valid for Figures 3-4, 6, 8-10. Specifically, $v^*_{T\infty}(\xi)$ and $v^*_{LT}(\xi) \; \forall \xi \in [0,1]$ are lower and upper bounds, respectively, for the Timoshenko dimensionless displacement $v^*(\xi)$ corresponding to any $\lambda \in \mathcal{S}$. An analogous observation holds for the Bernoulli-Euler beams. The closed-form expressions of the dimensionless transverse displacement $v^*_{T\infty}(\xi)$ are evaluated in Appendix A and are reported in Table 2 for the nano-beams considered in Figures 2-10. Finally, Table 3 reports the exact solutions $v^*(\xi)$ for $\alpha = 0.5$, $\beta = 4$, $\lambda = 0.2$ and $\xi = 0, 0.2, 0.4, 0.6, 0.8, 1$.

## 6. Conclusions

An innovative stress-driven elastic mixture has been developed for Timoshenko nano-beams. The model has been formulated by convexly combining local and nonlocal phases and does not suffer the limit behavior of the strain-driven mixture as the local fraction tends to zero. Both integral and differential formulations of the mixture stress-driven model have been provided and relevant procedures for obtaining closed-form solutions (reported in Appendix A for some simple structural schemes) have been illustrated. Finally, extensive numerical solutions have been presented for stubby beams. In agreement with experimental evidences [46], a stiffening structural response has been enlightened for increasing values of the scale parameter.

**Acknowledgment** - Financial supports from the Italian Ministry of Education, University and Research (MIUR) in the framework of the Project PRIN 2015 "COAN 5.50.16.01" - code 2015JW9NJT - and from the research program ReLUIS 2018 are gratefully acknowledged.

| Name of the nano-beam | | Boundary conditions | | | |
|---|---|---|---|---|---|
| *C-F* | (Fig. 2) | $v(0) = 0,$ | $\phi(0) = 0,$ | $M(L) = 0,$ | $T(L) = F$ |
| *C-q* | (Fig. 3) | $v(0) = 0,$ | $\phi(0) = 0,$ | $M(L) = 0,$ | $T(L) = 0$ |
| *SG-F* | (Fig. 4) | $v(0) = 0,$ | $M(0) = 0,$ | $\phi(L) = 0,$ | $T(L) = F$ |
| *SS-m* | (Fig. 5) | $v(0) = 0,$ | $M(0) = 0,$ | $v(L) = 0,$ | $M(L) = m$ |
| *SS-q* | (Fig. 6) | $v(0) = 0,$ | $M(0) = 0,$ | $v(L) = 0,$ | $M(L) = 0$ |
| *CS-m* | (Fig. 7) | $v(0) = 0,$ | $\phi(0) = 0,$ | $v(L) = 0,$ | $M(L) = m$ |
| *CS-q* | (Fig. 8) | $v(0) = 0,$ | $\phi(0) = 0,$ | $v(L) = 0$ | $M(L) = 0$ |
| *CG-F* | (Fig. 9) | $v(0) = 0,$ | $\phi(0) = 0,$ | $\phi(L) = 0,$ | $T(L) = F$ |
| *CC-q* | (Fig. 10) | $v(0) = 0,$ | $\phi(0) = 0,$ | $v(L) = 0$ | $\phi(L) = 0$ |

Table 1: Fundamental schemes with corresponding boundary conditions.

| Nano-beam | | $v_{T\infty}^*(\xi)$ |
|---|---|---|
| *C-F* | (Fig. 2) | $-\frac{1}{6}\xi^3\alpha + \frac{1}{2}\xi^2\alpha + \frac{\alpha\xi}{\beta}$ |
| *C-q* | (Fig. 3) | $\frac{1}{24}\xi^4\alpha - \frac{1}{6}\xi^3\alpha + \frac{1}{4}\xi^2\alpha + \frac{-\xi^2\alpha/2 + \xi\alpha}{\beta}$ |
| *SG-F* | (Fig. 4) | $-\frac{1}{6}\xi^3\alpha + \frac{1}{2}\xi\alpha + \frac{\alpha\xi}{\beta}$ |
| *SS-m* | (Fig. 5) | $\frac{1}{6}\xi^3\alpha - \frac{1}{6}\xi\alpha$ |
| *SS-q* | (Fig. 6) | $\frac{1}{24}\xi^4\alpha - \frac{1}{12}\xi^3\alpha + \frac{1}{24}\xi\alpha + \frac{-\xi^2\alpha + \xi\alpha}{2\beta}$ |
| *CS-m* | (Fig. 7) | $\frac{(\xi^3\alpha - \xi^2\alpha)\beta + 6\xi^2\alpha - 6\xi\alpha}{4\beta + 12}$ |
| *CS-q* | (Fig. 8) | $\alpha\frac{(2\xi^4 - 5\xi^3 + 3\xi^2)\beta^2 + 6(\xi^4 - 2\xi^3 - 4\xi^2 + 5\xi)\beta - 72\xi^2 + 72\xi}{48\beta^2 + 144\beta}$ |
| *CG-F* | (Fig. 9) | $-\frac{1}{6}\xi^3\alpha + \frac{1}{4}\xi^2\alpha + \frac{\alpha\xi}{\beta}$ |
| *CC-q* | (Fig. 10) | $\frac{1}{24}\xi^4\alpha - \frac{1}{12}\xi^3\alpha + \frac{1}{24}\xi^2\alpha + \frac{-\xi^2\alpha + \xi\alpha}{2\beta}$ |

Table 2: $v_{T\infty}^*(\xi)$ for different types of nano-beams.

| $\xi$ | C-F | C-q | SG-F | SS-m | SS-q | CS-m | CS-q | CG-F | CC-q |
|---|---|---|---|---|---|---|---|---|---|
| 0 | 0 | 0 | 0 | 0 | 0 | 0 | 0 | 0 | 0 |
| 0.2 | 0.056216 | 0.041956 | 0.130010 | -0.027542 | 0.021572 | -0.033140 | 0.018086 | 0.048079 | 0.015244 |
| 0.4 | 0.144422 | 0.090724 | 0.256206 | -0.051270 | 0.033961 | -0.058593 | 0.029400 | 0.110247 | 0.024375 |
| 0.6 | 0.256134 | 0.138857 | 0.367918 | -0.060514 | 0.033961 | -0.066719 | 0.030096 | 0.176925 | 0.024375 |
| 0.8 | 0.382330 | 0.181842 | 0.456125 | -0.046252 | 0.021572 | -0.049586 | 0.019496 | 0.239092 | 0.015244 |
| 1.0 | 0.512340 | 0.217551 | 0.512340 | 0 | 0 | 0 | 0 | 0.287172 | 0 |

Table 3: Exact solutions $v^*(\xi)$ for $\alpha = 0.5$, $\beta = 4$ and $\lambda = 0.2$.

## A. Appendix: Closed-form solutions for Timoshenko nano-beams

Next, closed-form solutions for Timoshenko nano-beams are provided by applying the integral procedure of Section 4.1 or, equivalently, the differential procedure of Section 4.2. For the most nano-beams considered in this work, the dimensionless transverse displacement $v^*$ may be expressed by the following formula:

$$v^*(\xi, \alpha, \beta, \lambda) = \mathcal{F}(\xi, \alpha, \lambda) + \frac{1}{\beta} \mathcal{H}(\xi, \alpha, \lambda), \tag{33}$$

or, equivalently, the transverse displacement may be represented as

$$v(x, \alpha, EI, \kappa GA, L, L_c, \mathcal{L}) = \frac{\mathcal{L}}{EI} \hat{\mathcal{F}}(x, \alpha, L, L_c) + \frac{\mathcal{L}}{\kappa GA} \hat{\mathcal{H}}(x, \alpha, L, L_c), \tag{34}$$

where $\mathcal{L}$ is one of the loads $m$, $F$ and $q$. Basically, the right-hand side of (33) or (34) is the sum of two addends: the first addend represents the flexural deflection and the second addend is the shear deformation. Formulae (33) and (34) occur not only in statically determinate (or isostatic) structures, such as *C-F*, *C-q*, *SG-F*, *SS-q* and *SS-m*, but also in some statically indeterminate (or hyperstatic) structures. In Table 2, these hyperstatic structures are *CG-F* and *CC-q*. In contrast, the dimensionless displacement $v^*$ of the *CS-m* and *CS-q* nano-beams can not be represented with formula (33) as the sum of two terms, one independent on $\beta$ and the other inversely proportional to $\beta$. In fact, the dimensionless displacement $v^*$ of the *CS-q* nano-beam exhibits the following limit (the closed-form expression of $v^*$ is too long and is not reported):

$$\lim_{\lambda \to +\infty} v^*(\xi, \alpha, \beta, \lambda) = \frac{\alpha}{48\,\beta^2 + 144\,\beta} \left[ \left(2\,\xi^4 - 5\,\xi^3 + 3\,\xi^2\right)\beta^2 + \right. $$
$$\left. + 6\left(\xi^4 - 2\,\xi^3 - 4\,\xi^2 + 5\,\xi\right)\beta - 72\,\xi^2 + 72\,\xi \right]. \tag{35}$$

See also the following subsections, where the dependence of $v^*$ on $\alpha$ and $\beta$ in the left-hand side of (33) is omitted for the sake of brevity.

### A.1. Cantilever subject to a concentrated force at the free end (C-F)

The closed-form expression of $v(x)$ is used in the following relation in order to obtain the dimensionless transverse displacement

$$v^*(\xi) = v(x)\frac{EI}{FL^3}, \tag{36}$$

which depends on the dimensionless parameters $\alpha$, $\beta$, $\lambda$ and $\xi$. Highlighting the dependence of $v^*(\xi)$ on $\lambda$, we rename $v^*(\xi)$ to $v^*(\xi, \lambda)$, helping us to present the following closed-form expressions of dimensionless displacements. The dimensionless transverse displacement of the cantilever is

$$v^*(\xi, \lambda) = \mathcal{F}_{CF}(\xi, \lambda) + \frac{1}{\beta}\mathcal{H}_{CF}(\xi, \lambda), \tag{37}$$

where $\mathcal{F}_{CF}(\xi, \lambda)$ has been found in [42] and $\mathcal{H}_{CF}(\xi, \lambda)$ is given by

$$\mathcal{H}_{CF}(\xi, \lambda) = \xi + \frac{1}{2}(1-\alpha)\left[\lambda\left(e^{-\xi\lambda^{-1}} + e^{-\lambda^{-1}} - e^{(\xi-1)\lambda^{-1}}\right) - \lambda\right]. \tag{38}$$

The maximum dimensionless transverse displacement of the cantilever is given by

$$v^*(1, \lambda) = \mathcal{F}_{CF}(1, \lambda) + \frac{1}{\beta}\mathcal{H}_{CF}(1, \lambda), \tag{39}$$

where

$$\mathcal{F}_{CF}(1, \lambda) = 1/3 + (\alpha - 1)\left(e^{-\lambda^{-1}}\lambda^3 - \lambda^3 + e^{-\lambda^{-1}}\lambda^2 + \lambda/2\right), \tag{40}$$

$$\mathcal{H}_{CF}(1, \lambda) = 1 + (1 - \alpha)\left(e^{-\lambda^{-1}}\lambda - \lambda\right). \tag{41}$$

Other significant dimensionless displacements are:

$$v^*_{T\infty}(\xi) = \lim_{\lambda \to +\infty} v^*(\xi, \lambda) = -\frac{1}{6}\xi^3\alpha + \frac{1}{2}\xi^2\alpha + \frac{\alpha\,\xi}{\beta}, \tag{42}$$

$$v^*_{LT}(\xi) = \lim_{\lambda \to 0^+} v^*(\xi, \lambda) = -\frac{1}{6}\xi^3 + \frac{1}{2}\xi^2 + \frac{\xi}{\beta}, \tag{43}$$

$$v^*_{LT}(1) = 1/3 + \beta^{-1}. \tag{44}$$

Entities $v^*_{T\infty}(\xi)$ in (42) and $v^*_{LT}(\xi)$ in (43) are defined in Section 5. $v^*_{T\infty}(\xi)$ is also reported in Table 2 for different types of nano-beams. The simple formula (44) allows explaining easily the use of the parameter $\beta$ in dimensional analysis for the transition from dimensionless quantities to dimensional quantities. In fact, multiplying both sides of (44) by $\frac{FL^3}{EI}$ gives

$$\frac{FL^3}{EI}v^*_{LT}(1) = \frac{FL^3}{3EI} + \frac{FL^3}{\beta EI}. \tag{45}$$

The transverse displacement of the local Timoshenko beam is $v_{LT}(x) = \frac{FL^3}{EI} v_{LT}^*(\xi)$. Moreover assuming $\kappa GA = \beta \frac{EI}{L^2}$, relation (45) becomes

$$v_{LT}(L) = \frac{FL^3}{3EI} + \frac{FL}{\kappa GA}, \tag{46}$$

which is the well-known formula of the maximum transverse displacement of a Timoshenko cantilever subject to a point load at the free end. The dimensional analysis for more complicated formulas such as (39) is done in a similar way.

### A.2. Cantilever subject to a uniformly distributed load (C-q)

Following the steps in Appendix A.1, the dimensionless transverse displacement is

$$v^*(\xi) = v(x) \frac{EI}{qL^4}, \tag{47}$$

which is renamed $v^*(\xi, \lambda)$. The maximum dimensionless transverse displacement is

$$v^*(1, \lambda) = \mathcal{F}_{Cq}(1, \lambda) + \frac{1}{\hat{\beta}} \mathcal{H}_{Cq}(1, \lambda), \tag{48}$$

where

$$\mathcal{F}_{Cq}(1, \lambda) = \frac{1}{8} + \frac{1}{4}(\alpha - 1)\left(e^{-\lambda^{-1}}\lambda^2 - \lambda^2 + \lambda\right), \tag{49}$$

$$\mathcal{H}_{Cq}(1, \lambda) = \frac{1}{2}\left[1 + (1 - \alpha)\left(e^{-\lambda^{-1}}\lambda - \lambda\right)\right]. \tag{50}$$

Notice that shear contribution $\mathcal{H}_{Cq}(1, \lambda)$ in (50) for cantilever with distributed load is equal to half of the shear contribution $\mathcal{H}_{CF}(1, \lambda)$ in (41) for cantilever with point load. Other significant solutions are

$$v_{T\infty}^*(\xi) = \lim_{\lambda \to +\infty} v^*(\xi,\lambda) = \frac{1}{24}\,\xi^4 \alpha - \frac{1}{6}\,\xi^3 \alpha + \frac{1}{4}\,\xi^2 \alpha + \frac{-\xi^2 \alpha/2 + \xi\alpha}{\beta}, \quad (51)$$

$$v_{LT}^*(\xi) = \lim_{\lambda \to 0^+} v^*(\xi,\lambda) = \frac{1}{24}\,\xi^4 - \frac{1}{6}\,\xi^3 + \frac{1}{4}\,\xi^2 + \frac{-\xi^2/2 + \xi}{\beta}, \quad (52)$$

$$v_{LT}(L) = \frac{qL^4}{EI}v_{LT}^*(1) = \frac{1}{8}\frac{qL^4}{EI} + \frac{1}{2}\frac{qL^2}{\kappa GA}. \quad (53)$$

*A.3. Simply supported beam with uniformly distributed load (SS-q)*

The maximum dimensionless transverse displacement is

$$v^*\left(\frac{1}{2},\lambda\right) = \mathcal{F}_{SSq}\left(\frac{1}{2},\lambda\right) + \frac{1}{\beta}\mathcal{H}_{SSq}\left(\frac{1}{2},\lambda\right), \quad (54)$$

where

$$\mathcal{F}_{SSq}\left(1/2,\lambda\right) =$$
$$5/384 + (\alpha - 1)\left(\left(\lambda^4 + \lambda^3/2\right)\left(e^{-(2\lambda)^{-1}} - e^{-\lambda^{-1}}/2 - 1/2\right) + \lambda^2/8\right), \quad (55)$$

$$\mathcal{H}_{SSq}(1/2,\lambda) =$$
$$1/8 + (1 - \alpha)\left(\left(\lambda^2 + \lambda/2\right)\left(e^{-(2\lambda)^{-1}} - e^{-\lambda^{-1}}/2\right) - \lambda^2/2 - \lambda/4\right), \quad (56)$$

Other significant solutions are

$$v_{T\infty}^*(\xi) = \lim_{\lambda \to +\infty} v^*(\xi, \lambda) = \frac{1}{24}\xi^4\alpha - \frac{1}{12}\xi^3\alpha + \frac{1}{24}\xi\alpha + \frac{-\xi^2\alpha/2 + \xi\alpha/2}{\beta},$$

(57)

$$v_{LT}^*(\xi) = \lim_{\lambda \to 0^+} v^*(\xi, \lambda) = \frac{1}{24}\xi^4 - \frac{1}{12}\xi^3 + \frac{1}{24}\xi + \frac{-\xi^2/2 + \xi/2}{\beta}, \qquad (58)$$

$$v_{LT}\left(\frac{L}{2}\right) = \frac{qL^4}{EI}v_{LT}^*\left(\frac{1}{2}\right) = \frac{5}{384}\frac{qL^4}{EI} + \frac{1}{8}\frac{qL^2}{\kappa GA}. \qquad (59)$$

*A.4. Nano-beam with a clamped end and a guided end subject to a point force (CG-F)*

Following the steps in Appendix A.1, the maximum dimensionless transverse displacement is

$$v^*(1, \lambda) = \mathcal{F}_{CGF}(1, \lambda) + \frac{1}{\beta}\mathcal{H}_{CGF}(1, \lambda), \qquad (60)$$

where

$$\mathcal{F}_{CGF}(1, \lambda) =$$

$$1/12 + (\alpha - 1)\left(e^{-\lambda^{-1}}\lambda^3 - \lambda^3 + e^{-\lambda^{-1}}\lambda^2 + e^{-\lambda^{-1}}\lambda/4 + \lambda/4\right), \qquad (61)$$

$$\mathcal{H}_{CGF}(1, \lambda) = 1 + (1 - \alpha)\left(e^{-\lambda^{-1}}\lambda - \lambda\right). \qquad (62)$$

Notice that the shear contribution $\mathcal{H}_{CGF}(1,\lambda)$ in (62) is equal to the shear contribution $\mathcal{H}_{CF}(1,\lambda)$ in (41). Other significant solutions are

$$v_{T\infty}^*(\xi) = \lim_{\lambda \to +\infty} v^*(\xi,\lambda) = -\frac{1}{6}\,\xi^3\alpha + \frac{1}{4}\,\xi^2\alpha + \frac{\alpha\,\xi}{\beta},\qquad(63)$$

$$v_{LT}^*(\xi) = \lim_{\lambda \to 0^+} v^*(\xi,\lambda) = -\frac{1}{6}\,\xi^3 + \frac{1}{4}\,\xi^2 + \frac{\xi}{\beta},\qquad(64)$$

$$v_{LT}(L) = \frac{FL^3}{EI}\,v_{LT}^*(1) = \frac{FL^3}{12EI} + \frac{FL}{\kappa GA}.\qquad(65)$$

**Acknowledgment** - Financial supports from the Italian Ministry of Education, University and Research (MIUR) in the framework of the Project PRIN 2015 "COAN 5.50.16.01" - code 2015JW9NJT - and from the research program ReLUIS 2018 are gratefully acknowledged.

# Figure

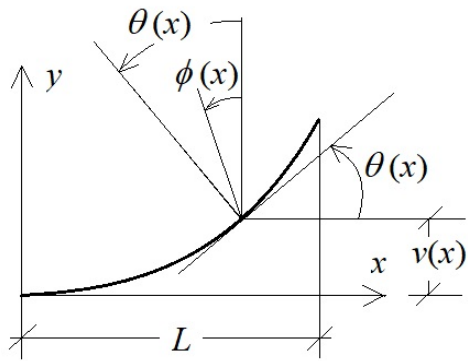

Figure 1: Kinematic variables defining the deformation of the beam.

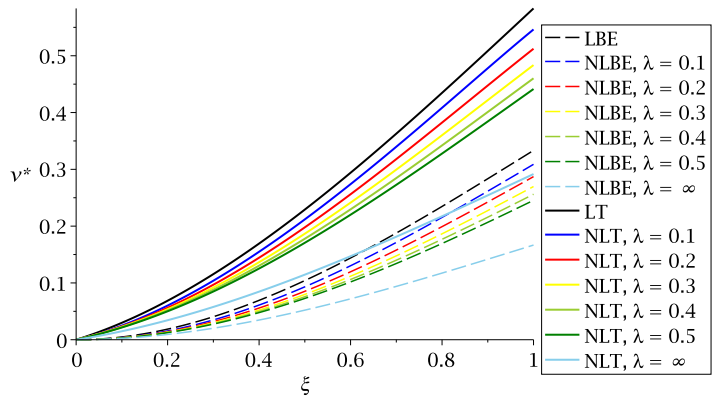

Figure 2: Cantilever with concentrated load at the free end ($C$-$F$): $v^*$ against $\xi$ for $\beta = 4$.

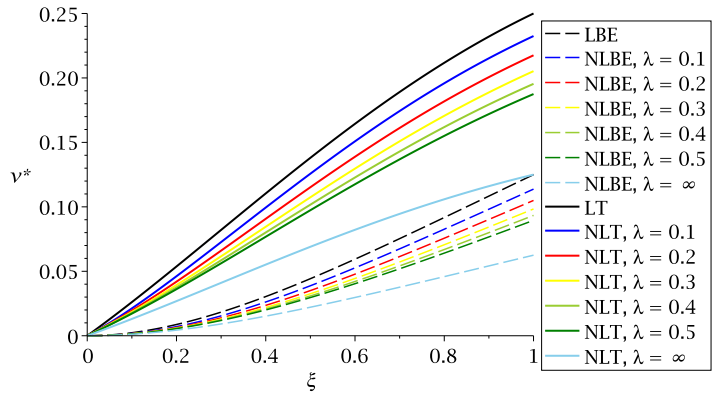

Figure 3: Cantilever with uniformly distributed load ($C$-$q$): $v^*$ against $\xi$ for $\beta = 4$.

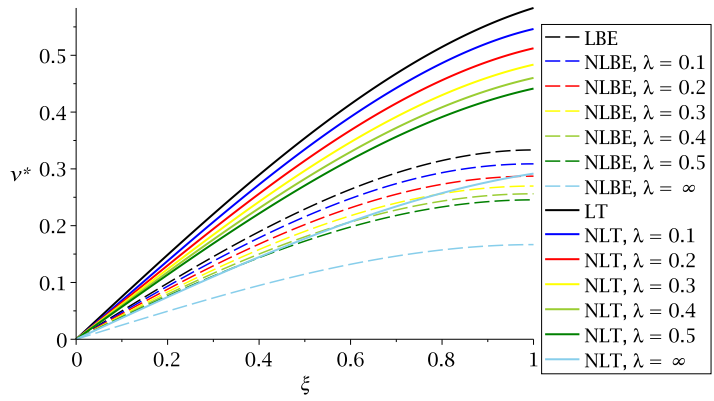

Figure 4: Beam with a simply supported end and a guided end subject to transverse concentrated load ($SG$-$F$): $v^*$ against $\xi$ for $\beta = 4$.

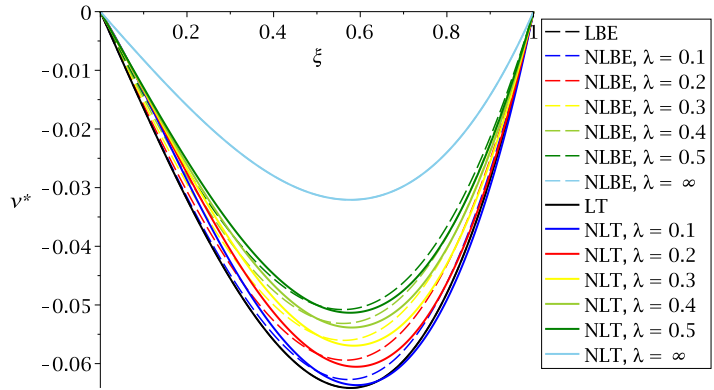

Figure 5: Simply supported beam with concentrated couple at the right end ($SS$-$m$): $v^*$ against $\xi$ for $\beta = 4$.

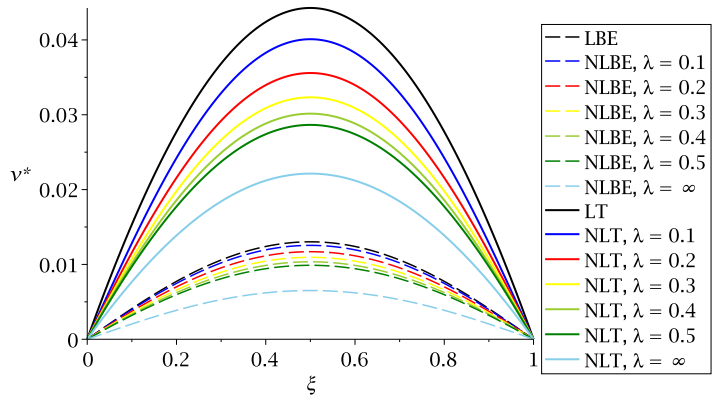

Figure 6: Simply supported beam with uniformly distributed load (*SS-q*): $v^*$ against $\xi$ for $\beta = 4$.

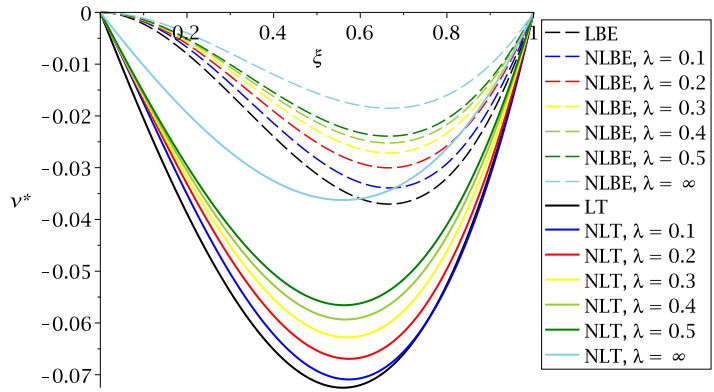

Figure 7: Beam with a clamped end and a simply supported end subject to a concentrated couple (*CS-m*): $v^*$ against $\xi$ for $\beta = 4$.

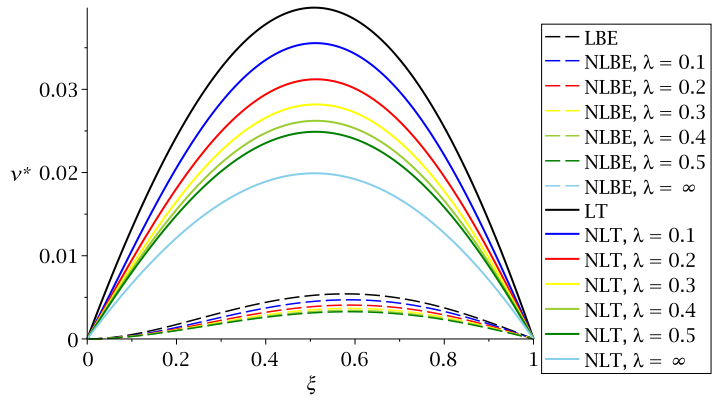

Figure 8: Beam with a clamped end and a simply supported end with uniformly distributed load ($CS$-$q$): $v^*$ against $\xi$ for $\beta = 4$.

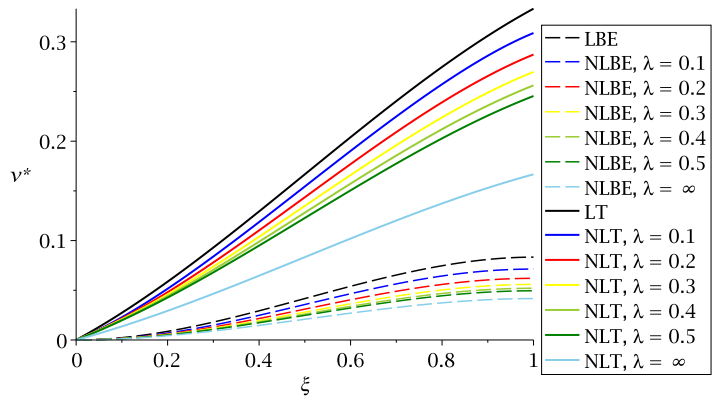

Figure 9: Beam with a clamped end and a guided end subject to transverse concentrated load ($CG$-$F$): $v^*$ against $\xi$ for $\beta = 4$.

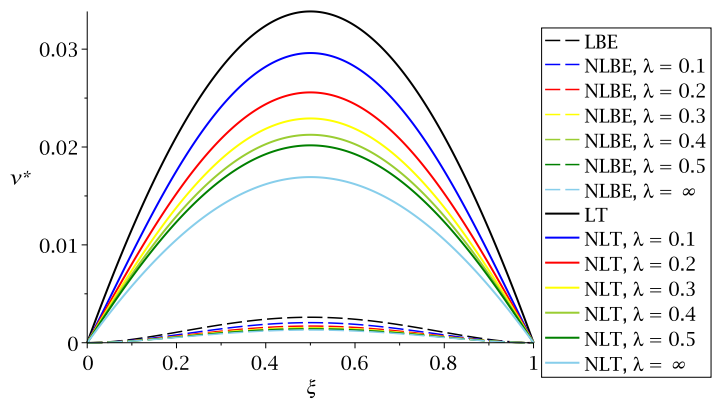

Figure 10: Fixed beam with uniformly distributed load ($CC$-$q$): $v^*$ against $\xi$ for $\beta = 4$.